%% file: main.tex
\documentclass[conference]{IEEEtran}
\IEEEoverridecommandlockouts

\input{preamble.tex}

\begin{document}

\bstctlcite{IEEEexample:BSTcontrol}

\title{Concatenated Codes for Recovery From\\ Multiple Reads of DNA Sequences \\
\thanks{A. Lenz and I. Maarouf  contributed equally to the results in this paper.

This project was  supported by the European Research Council (ERC) under the European Union’s Horizon 2020 research and innovation programme (grant agreement No. 801434), the Swedish Research Council under grant 2016-04253, and the TUM Global Visiting Professor program.
}
}

\author{\IEEEauthorblockN{{\bf Andreas Lenz}\IEEEauthorrefmark{1}, {\bf Issam Maarouf}\IEEEauthorrefmark{2}, {\bf Lorenz Welter}\IEEEauthorrefmark{1},\\ {\bf Antonia Wachter-Zeh}\IEEEauthorrefmark{1}, {\bf Eirik Rosnes}\IEEEauthorrefmark{2}, {\bf Alexandre Graell i Amat}\IEEEauthorrefmark{3}\IEEEauthorrefmark{2}}
\IEEEauthorblockA{
	\IEEEauthorrefmark{1}Institute for Communications Engineering, Technical University of Munich, DE-80333 Munich, Germany
}
\IEEEauthorblockA{
	\IEEEauthorrefmark{2}Simula UiB, N-5006  Bergen,  Norway
}
\IEEEauthorblockA{
	\IEEEauthorrefmark{3}Department of Electrical Engineering, Chalmers University of Technology, SE-41296 Gothenburg, Sweden
}}
\maketitle

\begin{abstract}
Decoding sequences that stem from multiple transmissions of a codeword over an insertion, deletion, and substitution channel is a critical component of efficient deoxyribonucleic acid (DNA) data storage systems. In this paper, we consider a concatenated coding scheme with an outer low-density parity-check code and either an inner convolutional code or a  block code. %
We propose two new decoding algorithms for inference from multiple received sequences, both combining the inner code and channel to a joint hidden Markov model to infer symbolwise a posteriori probabilities (APPs). The first decoder computes the exact APPs by jointly decoding the received sequences, whereas the second decoder approximates the APPs by combining the results of separately decoded received sequences. Using the proposed algorithms, we evaluate the performance of decoding multiple received sequences by means of achievable information rates and Monte-Carlo simulations. We show significant performance gains compared to a single received sequence. %

\end{abstract}

\input{introduction.tex}

\input{model.tex}

\input{multiple.tex}

\input{simulation.tex}

\input{conclusion.tex}

\bibliographystyle{IEEEtran}
\bibliography{etalabbr,refs}

\end{document}

%% file: preamble.tex
\usepackage{amsmath,amssymb,amsfonts,tikz,pgfplots,bm}
\usepackage{cite,booktabs,tabularx}
\usepackage{algorithmic}
\usepackage{graphicx}
\usepackage{textcomp}
\usepackage{xcolor}
\usepackage{verbatim}

\newcommand{\len}{\ensuremath{N}}
\newcommand{\leni}{\ensuremath{n}}
\newcommand{\leno}{\ensuremath{N_\mathsf{o}}}
\newcommand{\numS}{\ensuremath{M}}
\newcommand{\bl}{\leno+m}

\newcommand{\D}{\mathsf{D}}
\newcommand{\I}{\mathsf{I}}
\newcommand{\T}{\mathsf{T}}
\renewcommand{\S}{\mathsf{S}}

\renewcommand{\d}{\ensuremath{\boldsymbol{d}}}
\newcommand{\s}{\ensuremath{\boldsymbol{s}}}
\renewcommand{\u}{\ensuremath{\boldsymbol{u}}}
\newcommand{\w}{\ensuremath{\boldsymbol{w}}}
\newcommand{\x}{\ensuremath{\boldsymbol{x}}}
\newcommand{\y}{\ensuremath{\boldsymbol{y}}}
\renewcommand{\v}{\ensuremath{\boldsymbol{v}}}

\newcommand{\Al}{\ensuremath{\Sigma}}
\newcommand{\F}{\ensuremath{\mathbb{F}}}

\usetikzlibrary{positioning,calc}
\tikzstyle{block} = [rectangle, draw, text centered, rounded corners, minimum height=1.5em,fill=black!5!]

\def\approxprop{%
	\def\p{%
		\setbox0=\vbox{\hbox{$\propto$}}%
		\ht0=0.6ex \box0 }%
	\def\s{%
		\vbox{\hbox{$\sim$}}%
	}%
	\mathrel{\raisebox{0.7ex}{%
			\mbox{$\underset{\s}{\p}$}%
	}}%
}

\renewcommand{\baselinestretch}{0.96} 

%% file: introduction.tex
\section{Introduction}
Coding-theoretic aspects of data storage in deoxyribonucleic acid (DNA) have recently gained significant attention. This new line of research has been triggered by a manifold of successful experiments %
\cite{yazdi_portable_2017,organick_random_2018,chandak_overcoming_2020}
that demonstrated the viability of DNA as a reliable medium for archival data storage. Following the pioneering experiments, a variety of related information-theoretic problems with different foci have been identified. One of these problems is the reliable transmission over channels that introduce insertions, deletions, and substitutions (IDSs)~\cite{heckel_characterization_2019}, motivated by the fact that the sequencing process often introduces errors in the form of IDSs.

There is a variety of literature dedicated to the study of error-correcting codes for the IDS channel. While the list of related papers is long, the line of research most relevant to our work is the one inspired by Davey and MacKay \cite{davey_reliable_2001}. In their paper, they propose a class of concatenated codes together with a decoding algorithm that represents the inner code and the channel by a hidden Markov model (HMM), which allows inferring a posteriori probabilities that are passed to the outer decoder. Following this work, the Markov process of convolutional codes was incorporated into that of the IDS channel, allowing to run standard decoding algorithms of convolutional codes for the IDS channel \cite{mansour_convolutional_2010,buttigieg_improved_2015}. The approach of \cite{davey_reliable_2001} was later refined with an improved decoding algorithm \cite{briffa_improved_2010} and using inner marker codes \cite{inoue_adaptive_2012}. Recently, it was shown that synchronization is also possible without an inner code using a standalone low-density parity-check (LDPC) code~\cite{shibata_design_2019}.

\newcommand{\skiplength}{ }
The vast majority of previous studies has focused on the case of error correction after a single transmission over an IDS channel.
In DNA-based storage, however, data is typically synthesized into many short DNA strands, and each of these strands is replicated thousands of times. Therefore, when accessing the stored data via DNA sequencing, multiple copies of each original strand are read.   With this in mind, in this paper we deviate from previous work and consider error-correcting codes for multiple transmissions of a coded sequence over a channel that is impaired by IDS errors. In particular, we propose two new decoding algorithms for multiple received sequences and show that significant gains can be achieved as compared to a single transmission. The first decoding algorithm, mainly a benchmark for the second decoder, is an optimal symbolwise a posteriori probability (APP) decoder. While this algorithm has exponential complexity in the number of received sequences, we show that it is possible to achieve competitive performance with a second, practical decoder that has only linear complexity. For our results we employ a new concatenated coding scheme with an outer LDPC code and an inner convolutional code and compare with the watermark codes from \cite{davey_reliable_2001}. To the best of our knowledge, this is the first work to propose decoding algorithms for the case of coded multiple transmission and to evaluate coding schemes as well as achievable information rates (AIRs) over this channel.

Related to our work is the study of sequence reconstruction,  introduced by Levenshtein \cite{levenshtein_efficient_2001}. Here, a sequence is repeatedly transmitted over an erroneous adversarial channel and the goal is to analytically quantify, as a function of the sequence length, the number of sequences  that are required to guarantee correct reconstruction of the transmitted sequence with zero error probability. Levenshtein's initial work on insertion and deletion errors has recently been extended by additionally allowing substitution errors \cite{sini_reconstruction_2019}.

Similar in spirit to our work, yet different in methods and objectives, is the research of trace reconstruction over deletion channels \cite{brakensiek_coded_2019,cheraghchi_coded_2020}, which is the probabilistic variant of the sequence reconstruction problem. While these works mainly derive asymptotic results, here we discuss a fixed number of short, finite-length received sequences over a channel that additionally allows insertions and substitutions. More recently, the trace reconstruction problem has also been formulated for a fixed number of sequences with a larger focus on algorithmic aspects \cite{srinivasavaradhan_symbolwise_2019,sabary_reconstruction_2020}. However, these works consider only uncoded  sequences, while we are interested in the case of coded transmission.  Finally, in \cite{abroshan_coding_2019} Varshamov-Tenegolts codes have been proposed for error correction over multiple channels that introduce a fixed number of deletions.

%% file: model.tex
\section{System Model} \label{sec:model}
\subsection{Channel Model}
\begin{figure}
	\centering
	\includegraphics{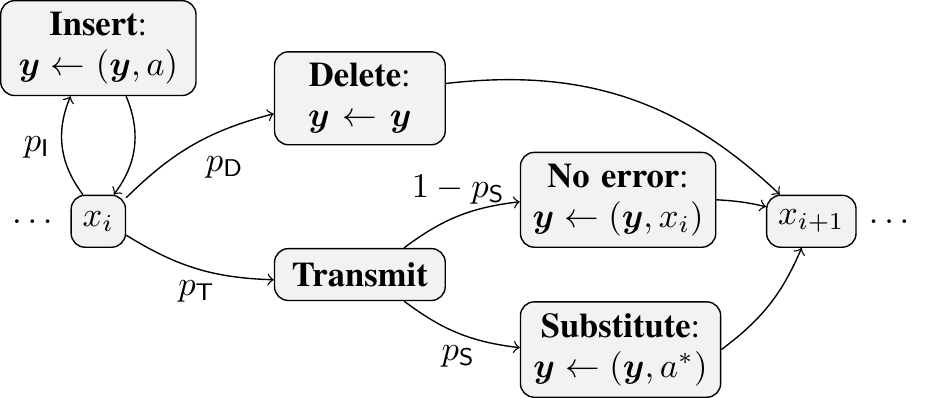}
	
	\caption{State-based IDS channel \cite{davey_reliable_2001}. }
	\label{fig:ids:channel}
\end{figure}

We consider a state-based representation of the IDS process \cite{zigangirov_sequential_1969,davey_reliable_2001} as depicted in Fig.~\ref{fig:ids:channel}.
Let $\x = (x_1,\dots,x_{\len})$, $x_i \in \Al_q = \{0,1,\dots,q-1\}$ be the sequence to be transmitted over the channel. The transmitted symbols $x_i$ are successively queued for transmission and the received sequence $\y$ is assembled by the following procedure. When $x_i$ enters the queue, the channel enters state $x_i$ and there are three different possible events. First, a uniformly random symbol $a \in \Al_q$ is inserted to the received sequence $\y$ with probability $p_\I$. In this case, $x_i$ remains in the queue and the channel returns to state $x_i$. Second, the symbol $x_i$ is deleted with probability $p_\D$ without appending it to $\y$ and the next symbol $x_{i+1}$ is enqueued. Third, $x_i$ is transmitted with probability $p_\T = 1-p_\I -p_\D$. In this case, the symbol is substituted with a random symbol $a^* \neq x_i$ with probability $p_\S$ and the next transmit symbol $x_{i+1}$ is enqueued. When the last transmit symbol $x_{\len}$ leaves the queue, the procedure finishes and the channel outputs $\y$. Note that $\y = (y_1,\dots,y_{\len'})$ has random length $\len'$, depending on the channel realization. 
\subsection{Forward Error-Correction Scheme} \label{subsec:fec:scheme}

In this work, we will use a concatenation of an outer code with an inner code as our forward error-correction scheme, following \cite{davey_reliable_2001}. 
The role of the inner code is to maintain synchronization and provide reliable soft information to the outer code, while the role of the outer code is to use this soft information to perform an accurate estimate of the transmitted information. In particular, given that the inner code provides reliable synchronization information, the outer code corrects the remaining errors on the synchronized sequence. In our scheme, the inner code is either a convolutional code as used in \cite{mercier_convolutional_2009}, or a watermark code as used in \cite{davey_reliable_2001}. Note that the watermark code can be thought of as a convolutional code with memory equal to zero, and we will therefore restrict some passages to convolutional inner codes only.
The outer code is either a binary or a nonbinary LDPC code. 

The information $\u=(u_1,\dots,u_K)$, $u_i \in \F_{2^k}$ of length $K$ is encoded by an $[\leno,K]_{2^k}$ outer code of length $\leno$ over the field $\F_{2^k}$. The rate of the outer code is $R_\mathsf{o} = \frac{K}{\leno}$. The resulting codeword $\w = (w_1,\dots,w_{\leno})$, $w_i \in \F_{2^k}$ is then encoded via either an $[\leni,k,m]_q$ inner convolutional code of blocklength $\leni$, binary input length $k$, memory $m$, and output alphabet $\Al_q$ or an $[\leni,k]_q$ inner block code. With this concatenation, each outer codeword symbol $w_i \in \F_{2^k}$ is mapped to $n$ symbols over $\Al_q$. We terminate the convolutional code, resulting in the codeword $\v = (v_1,\dots,v_{N})$, $v_i \in \Al_q$ of length $\len = (\bl) n$. The inner code consequently has rate $R_\mathsf{I}=\frac{\leno k}{\len \log q}$, and thus we transmit at an overall rate $R = R_\mathsf{o}R_\mathsf{I} = \frac{Kk}{N \log q}$. Before entering the IDS channel, the codeword $\v$ is offset by a pseudo-random sequence resulting in the final output $\x = (x_1,\dots,x_{N})$, $x_i \in \Al_q$. This offset sequence is known to the inner decoder, and is used to maintain synchronization at high probabilities of insertions or  deletions. According to the nomenclature in \cite{davey_reliable_2001} we call the combination of the inner block code and offset \emph{watermark} code.

\begin{figure}
	\centering
	\includegraphics{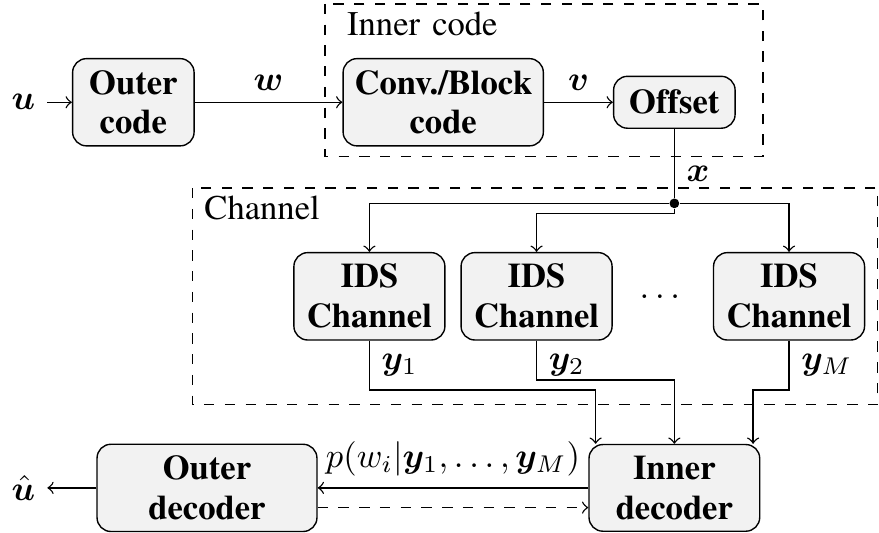}
	
	\caption{Communication via multiple transmissions over an IDS channel with a concatenated coding scheme. The IDS channel, depicted in Fig.~\ref{fig:ids:channel}, is fed $\numS$ times with the encoded transmit sequence $\x$. }
	\label{fig:system:model}
	\vspace{-2ex}
\end{figure}

The output sequence $\x$ is then transmitted $\numS$ times independently over an IDS channel resulting in received sequences $\y_1,\dots,\y_{\numS}$ corresponding to $\numS$ reads of the original strand. The inner decoder uses these received sequences to infer likelihoods for the symbols in $\w$. These likelihoods are then fed to the outer decoder, which decides on the decoded sequence $\hat{\u}$. Furthermore, we can also iterate between the inner and outer decoder, exchanging extrinsic information between them. The communication system is depicted in Fig.~\ref{fig:system:model}.    

%% file: multiple.tex
\section{Symbolwise A Posteriori Decoding of Insertion/Deletion/Substitution Channels} \label{sec:decoding}
One of the main challenges for decoding from IDS channels is the random memory associated with the insertion and deletion process. This is visualized by the receiver's inability to directly identify the origin of a received symbol $y_i$. Since insertions or deletions before symbol $y_i$ might have moved the symbol right or left in the received sequence, $y_i$ could be the result of transmitting a symbol $x_{i'}$ with $i'\neq i$. In the following, we describe how it is still possible to infer a posteriori likelihoods from this channel using a hidden Markov representation of the channel \cite{davey_reliable_2001,briffa_improved_2010,buttigieg_improved_2015}. 
We present and discuss this HMM for $\numS = 1$ first and extend it to $\numS>1$  multiple received sequences afterward. Here we restrict the discussion to convolutional inner codes, as the watermark code can be viewed as a convolutional code with $m=0$.
\subsection{Single Received Sequence}
The APPs of the outer codeword symbols are given by
$$ p(w_i|\y) = \frac{p(\y,w_i)}{p(\y)}~. $$
The standard approach to compute  $p(\y,w_i)$ for convolutionally encoded transmit sequences is to de-marginalize the distribution $p(\y,w_i)$ with respect to the memory states of the convolutional code  that correspond to the information symbol $w_i$ and then exploit the Markov property of the convolutional code. That is, given the memory state at time $i$, the output after time $i$ becomes independent of earlier states and the preceding output symbols. However, for the case of the IDS channel, this Markov property is violated through the memory of the channel. Therefore, it is useful to add a new hidden state variable, the so-called \emph{drift}\cite{davey_reliable_2001}. The drift $d_i$, $i =0,1,\dots,(\bl)$ is defined as the number of insertion events minus the number of deletion events that have occurred before symbol $x_{i\leni +1}$ is enqueued. Thus, by definition, $d_0 = 0$ and $d_{\bl} = N'-N$, both known to the receiver. In the resulting HMM, a transition from time $i-1$ to $i$ corresponds to a transmission of symbols $\x_{(i-1)n+1}^{in}$, where $\x_a^b = (x_a,x_{a+1},\dots x_b)$. Further, when transitioning from state $d_{i-1}$ to $d_{i}$ the HMM emits $n+d_{i}-d_{i-1}$ output symbols depending on both the previous and new drift.  The key property of the drift is that, by its inclusion as a new state variable inside the Markov process, the Markov property is restored. This is because the drift sequence itself forms a Markov chain and, conditioned on the memory state and drift at time $i$, the output of the channel after time $i$ becomes independent of previous memory states and drifts. Denoting the state variables of the convolutional code  by $s_i$, $i=0,1,\dots,(\bl)$, and introducing the joint state variable $\sigma_i = (s_i,d_i)$, we obtain by slight abuse of notation
\begin{align*}
	p(\y,w_i) = \sum_{(\sigma,\sigma'):w_i} p(\y,\sigma,\sigma')~,
\end{align*}
where the summation is over all state transition pairs $(\sigma,\sigma')$ that correspond to the information symbol $w_i$. Further, due to the Markov property, we can expand the joint probability to
\begin{align*}
	p&(\y,\!\sigma\!,\!\sigma') \!=\! p\!\left(\!\y_{1}^{(i\!-\!1)n+d}, \sigma\!\right)\!p\!\left(\!\y_{(i\!-\!1)n\!+d+\!1}^{in+d'}, \sigma'\big|\sigma\!\right)\!p\!\left(\!\y_{in\!+\!d'\!+1}^{\len'}\Big| \sigma'\!\right)\!.
\end{align*}
Abbreviating the above terms by $\alpha_{i-1}(\sigma)$, $\gamma_i(\sigma,\sigma')$, and $\beta_i(\sigma')$ in order of appearance, one can deduce the  forward and backward recursions
\begin{align*}
	\alpha_i(\sigma') &= \sum_{\sigma}\alpha_{i-1}(\sigma) \gamma_{i}(\sigma,\sigma')~,  \\
	\beta_i(\sigma) &= \sum_{\sigma'}\beta_{i+1}(\sigma') \gamma_{i+1}(\sigma,\sigma')~,
\end{align*}
with appropriate termination conditions incorporating the termination of the convolutional code and drift states.
The branch metric {$\gamma_i(\sigma,\sigma')=p(w_i)p(\y_{(i-1)n+d+1}^{in+d'}, d'\big|d,s,s'\!)$} can be efficiently computed using a lattice structure, see, e.g., \cite{bahl_decoding_1975,buttigieg_improved_2015}\footnote{Note that the lattice-based computation of the transition probability in \cite{bahl_decoding_1975,buttigieg_improved_2015} yields the joint conditional probability of $\y$ and drift $d'$ given a  transmit sequence. The generalization to nonbinary alphabets is straightforward. } using the transmit sequence corresponding to the state transition $s\rightarrow s'$. Having these expressions in hand it is possible to compute the APPs $p(w_i|\y)$ using the BCJR algorithm~\cite{bahl_optimal_1974}. Note that while the introduction of the drift state variable allows to exactly compute the APPs, it comes at the cost of additional complexity in terms of a larger number of states and state transitions, as discussed in Section~\ref{subsec:complexity}.
\subsection{Multiple Received Sequences}
In this work we consider the following two approaches of computing the a posteriori distribution $p(w_i|\y_1,\dots,\y_{\numS})$ for multiple sequences.

{\bf Joint decoding: }
Similar to the approach for a single received sequence, it is possible to introduce a drift state sequence  $\d_1,\dots,\d_{\numS}$, $\d_j = (d_{j,0},\dots,d_{j,\bl})$ for each of the received sequences. The resulting HMM has the combined $(\numS+1)$-dimensional state variables \mbox{$\sigma_i = (s_i,d_{1,i},\dots,d_{\numS,i} )$} with corresponding branch metric
\begin{align*}
	\gamma_i(\sigma,\sigma') = p(w_i) \prod_{j=1}^{\numS} p\left((\y_j)_{(i-1)n+d_j+1}^{in+d_j'}, d_j'\big|d_j,s,s'\!\right).
\end{align*}
The forward, backward recursion, and the terminal conditions are according to the single sequence case.

{\bf Separate decoding: }
Instead of computing the a posteriori likelihoods precisely, we now propose a different approach that allows to approximate the joint a posteriori likelihood using the marginal likelihoods $p(w_i|\y_j)$ as
$$ p(w_i|\y_1,\dots,\y_{\numS}) \approxprop \frac{\prod_{j=1}^{\numS} p(w_i|\y_j)}{p(w_i)^{\numS-1}}~. $$
The approximation is motivated by the case of memoryless channels, where the proportionality holds exactly. The individual APPs $p(w_i|\y_j)$ can be computed as in the case of a single received sequence by means of a BCJR decoder. While the approximation comes at reduced complexity as compared to the joint decoding approach, see Section \ref{subsec:complexity}, it also results in a loss of performance, which we will quantify in Section \ref{sec:results}. Note that it is also possible to separate the joint a posteriori likelihoods at other stages, such as directly between the channel and inner code or after decoding the LDPC code. However, we observed that for the considered parameters the proposed stage yields the best results. 
\subsection{Complexity Considerations} \label{subsec:complexity}
The number of operations performed by the inner BCJR decoder is proportional to the total number of edges of the trellis that is inferred from the HMM. In order to reduce its computational complexity, the drift states are usually limited to a fixed interval $d_i \in [d_\mathrm{min},d_{\max}]$ and the maximum number of insertions per symbol is limited to $I_\mathrm{max}$ \cite{davey_reliable_2001}. Denote by $\Delta = d_\mathrm{max}-d_{\min}+1$ the total number of drift states and by $\delta = n(I_\mathrm{max}+1)+1$ the number of possible drift transitions. 
The complexity, measured by the total number of edges in the trellis, of joint decoding is then given by
$$ C_{\mathrm{joint}} = \frac{N}{n} 2^{mk+k} (\Delta \delta)^\numS ~, $$
while the complexity of separate decoding is simply given by $\numS$ times the complexity of decoding a single sequence,
$$ C_{\mathrm{sep}} = \frac{N}{n} 2^{mk+k} \numS \Delta \delta ~. $$
Note that here we neglect the effect of trellis termination on the number of edges. Typical values of $\Delta \delta$ for the DNA storage application can easily be as high as $\Delta \delta \approx 1000$. This means that decoding the inner code is the main contributor to the decoding complexity and we thus neglect the complexity of the outer decoder in the following. It further becomes evident that the joint decoding approach only works for a very small number of sequences $\numS$ due to its exponential complexity.

%% file: simulation.tex
\section{Simulation Results} \label{sec:results}
In our simulations, we compare the two following set-ups. First, an inner convolutional code with an outer WiMAX binary LDPC code (CC + B-LDPC) and, second, an inner watermark code with an outer WiMAX-like nonbinary LDPC code (WM + NB-LDPC). Their performance is compared over a single and multiple IDS channels. Recall that in both schemes a pseudo-random offset sequence is used to improve the synchronization capability.
\subsection{Simulation Parameters}
Our simulations are done over an alphabet of size $q=4$, which equates to the four bases  $\{ \mathsf{A}, \mathsf{C}, \mathsf{G}, \mathsf{T} \}$ used in DNA and we set $p_\S = 0$ and $p_\I=p_\D$ throughout. The code parameters are summarized in Table \ref{tab:parameters}.
The outer LDPC code is decoded via belief propagation with a maximum number of $100$ iterations. %
Regarding the inner code, we set $I_\mathrm{max} = 2$ and the limit of the drift random variable in decoding is  set to be five times the standard deviation of the final drift at position $\len$, i.e., $d_{\max} = -d_\mathrm{min}  = 5 \sqrt{\vphantom{A}\smash{N \frac{p_\D}{1-p_\D}}}$. For multiple transmissions, we use $\numS = 2$, $5$, $10$, and $20$. 
\begin{table} 
	\setlength{\tabcolsep}{1.7pt}
	\begin{center}
		\caption{Code parameters}
		\vspace{-2ex}
		{\renewcommand{\arraystretch}{1.1}
			\begin{tabular}{cllcc} \specialrule{1.2pt}{0pt}{0pt}
				Scheme & \multicolumn{1}{c}{Inner Code} & \multicolumn{1}{c}{Outer Code} & Length & Rate \\ \specialrule{.8pt}{0pt}{0pt}
				CC+B-LDPC & $[1,1,2]_4$ Conv. & $[960,480]_2$ WiMAX & $962$ & $0.249$ \\
				WM+NB-LDPC & $[3,3]_4$ Waterm. & $[336,168]_{2^3}$ WiMAX-like & $1008$ & $0.25$ \\
				\specialrule{1.2pt}{0pt}{4pt}
			\end{tabular}\label{tab:parameters}}
	\end{center}
	The $[1,1,2]_4$ convolutional code is a binary convolutional code with generator polynomials $(1+D^2,1+D+D^2)$, where each two bits are combined to one DNA symbol. The $[3,3]_4$ watermark code is constructed as in \cite{davey_reliable_2001}.
	\vspace{-.4cm}
\end{table}
\subsection{Discussion of Numerical Results}
\begin{figure}
	\includegraphics{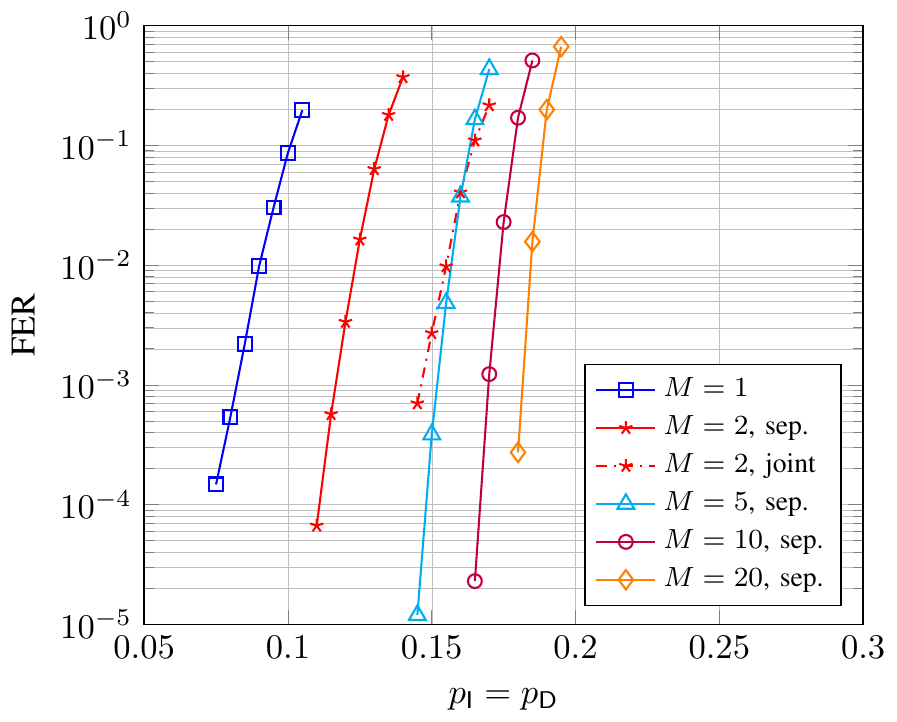}
	\vspace{-.55cm}
	\caption{FER performance versus $p_\I=p_\D$ of the CC + B-LDPC scheme for different number of transmissions $\numS$.}
    \label{CC + B-LDPC}
\end{figure}

\begin{figure}
	\includegraphics{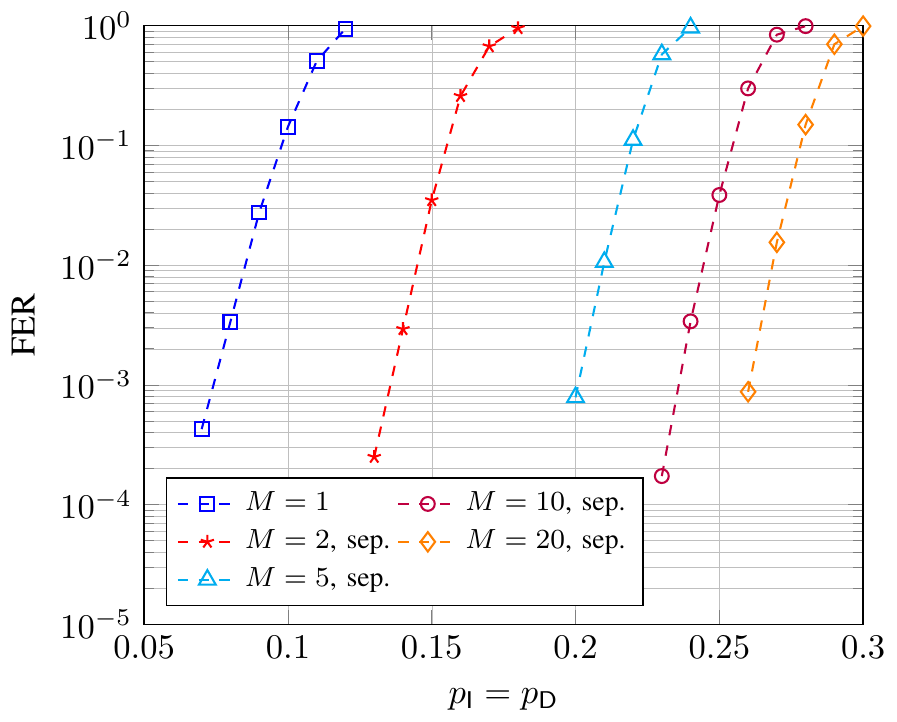}

	\vspace{-.15cm}
	\caption{FER performance versus $p_\I=p_\D$ of the WM+NB-LDPC scheme for different number of transmissions $\numS$.}
    \label{WM + NB-LDPC}
\vspace{-.3cm}
\end{figure}

\begin{figure}
	\includegraphics{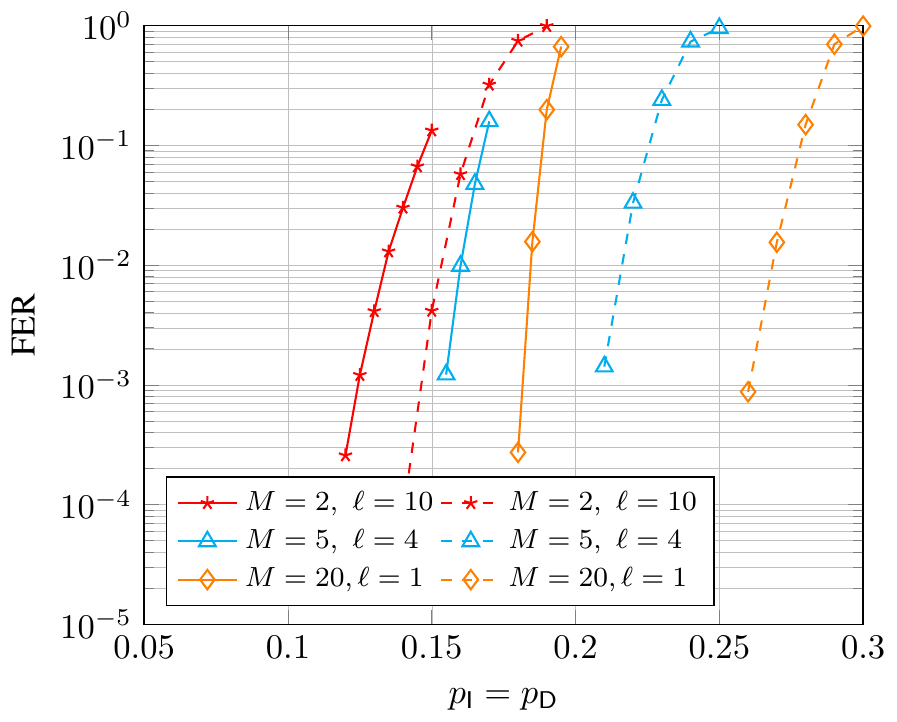}
	\vspace{-.65cm}
	\caption{FER performance versus $p_\I=p_\D$ of the CC + B-LDPC (solid) and WM + NB-LDPC (dashed) schemes with $M$ transmissions and $\ell$ iterations between inner and outer decoders.}
	\label{Sep. dec. + It. dec.}
	\vspace{-.4cm}
\end{figure}

\begin{figure}
	\includegraphics{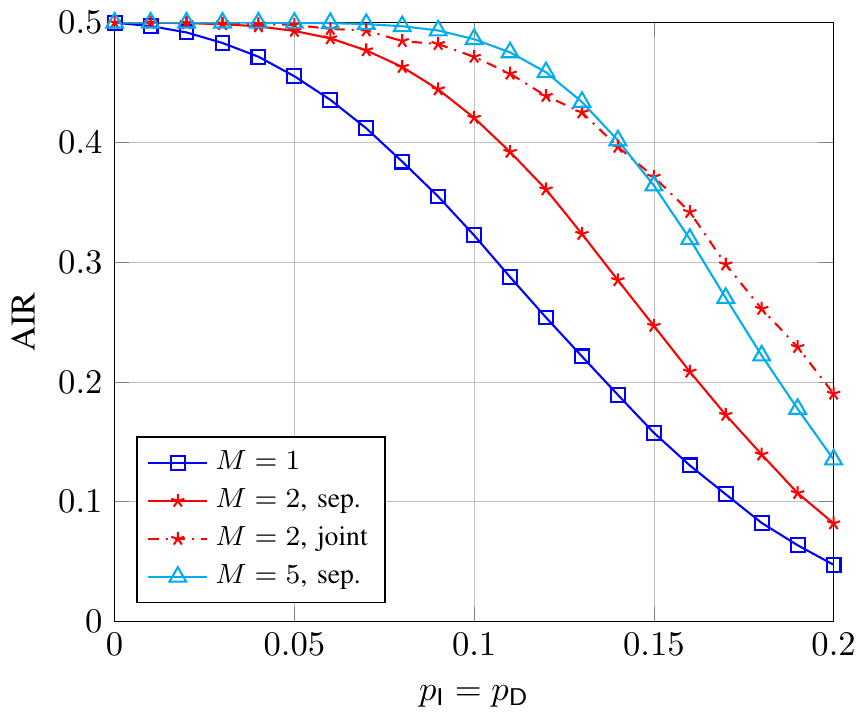}
	\vspace{-.55cm}
	\caption{AIRs versus $p_\I=p_\D$ using an inner convolutional code with multiple transmissions.}
	\label{fig:AIRs}
	
\end{figure}

Fig.~\ref{CC + B-LDPC} shows the frame error rate (FER) of the CC + B-LDPC scheme for single and multiple transmission over an IDS channel, while Fig.~\ref{WM + NB-LDPC} shows that of the WM + NB-LDPC scheme. Interestingly, when transmitting over a single IDS channel, the CC + B-LDPC scheme outperforms the WM + NB-LDPC scheme. However, for transmission over multiple IDS channels, this behavior reverses, and the WM + NB-LDPC scheme has superior performance for a larger number of sequences. While this result seems surprising at the first glance, there is a natural explanation for this phenomenon as follows. First notice that with the inner convolutional code, bit-level APPs $p(w_i|\y_j)$ with $w_i \in \F_2$ are computed, while the watermark code works over nonbinary inputs $w_i \in \F_{2^3}$. Now, for larger error probabilities $p_\I$, $p_\D$, the memory of the channel becomes more apparent and thus also the a posteriori distribution has increased correlation between different input symbols $w_i$. While this correlation is marginalized for the binary APPs, it is better captured by the nonbinary input symbols and thus the WM + NB-LDPC scheme achieves better performance for larger number of sequences or higher error probabilities. %
In terms of complexity, both schemes have the same $\Delta$ and $\numS$, but different $n$, $k$, $m$, and $\delta$. %
Comparing both complexities for equal blocklengths, we get \mbox{$C_{\mathrm{sep}}^{\text{conv.}}/C_{\mathrm{sep}}^{\text{watermark}} \approx 1.2.$} Thus, with these parameters the convolutional code is slightly more complex than the watermark code. Since the complexity increases exponentially with the memory $m$, larger memory convolutional codes are expected to become even more complex as compared to the watermark code. The same conclusion can be made when multiple channel transmissions are combined with inner-outer iterative decoding, since doing $\numS$ iterations has the same complexity as separate decoding of $\numS$ received sequences. Fig. \ref{Sep. dec. + It. dec.} shows the FER of both schemes when iterating $\ell$ times between the inner and outer decoder. We fix the total complexity such that the product of the number of iterations and the number of sequences  is equal to $20$. It is observed that the iterative scheme benefits more from a larger number of received sequences, however in practice this comes at the cost of additional sequencing costs. Finally, AIRs with an inner convolutional code are displayed in Fig.~\ref{fig:AIRs}. In agreement with our previous results, the achievable rates are significantly larger for multiple transmissions as compared to a single transmission. Not surprisingly, the joint decoder achieves higher rates. It exhibits a similar performance with $\numS=2$ transmissions as compared to the separate decoder with $\numS=5$ sequences.

\pgfplotsset{
	every axis plot/.append style={line width=.6pt},
	every axis plot post/.append style={
		every mark/.append style={line width=.6pt}
	}
}

%% file: conclusion.tex
\section{Future Work}
In view of the huge number of trellis states that the joint decoding approach entails, an immediate idea is to consider suboptimal decoding of the inner code, such as sequential decoding. There are however some obstacles that have to be overcome, including a huge number of edges per node and an appropriate metric. This will be addressed in future work.